\documentclass[journal,twoside]{IEEEtran}

\usepackage{etoolbox}
\makeatletter
\@ifundefined{color@begingroup}%
{\let\color@begingroup\relax
	\let\color@endgroup\relax}{}%
\def\fix@ieeecolor@hbox#1{%
	\hbox{\color@begingroup#1\color@endgroup}}
\patchcmd\@makecaption{\hbox}{\fix@ieeecolor@hbox}{}{\FAILED}
\patchcmd\@makecaption{\hbox}{\fix@ieeecolor@hbox}{}{\FAILED}

\usepackage{cite}
\usepackage{amsmath,amssymb,amsfonts}
\usepackage{graphicx}
\usepackage{multirow}
\usepackage{color}
\usepackage{soul}
\usepackage{pifont}
\usepackage[switch]{lineno}

\usepackage{algorithmic}
\usepackage{placeins}
\usepackage{stfloats}
\usepackage{fixltx2e}
\usepackage{lineno}

\usepackage{bm}
\begin{document}
	\title{\Huge Sequential-Scanning Dual-Energy CT Imaging Using High Temporal Resolution Image Reconstruction and Error-Compensated Material Basis Image Generation}
	\author{Qiaoxin Li, Ruifeng Chen, Peng Wang, Guotao Quan, Yanfeng Du, Dong Liang, Yinsheng Li
		\thanks{\emph{Corresponding authors: Yinsheng Li and Dong Liang.}}
		\thanks{Qiaoxin Li (e-mail: joosenli@outlook.com) is with Research Center for Medical Artificial Intelligence, Shenzhen Institutes of Advanced Technology, Chinese Academy of Sciences, Shenzhen, China and University of Chinese Academy of Sciences, Beijing, China.}		
		\thanks{Ruifeng Chen (e-mail: rf.chen1@siat.ac.cn) is with Research Center for Medical Artificial Intelligence, Shenzhen Institutes of Advanced Technology, Chinese Academy of Sciences, Shenzhen, China and with Laboratory of Image Science and Technology, Key Laboratory of Computer Network and Information Integration, Southeast University, Nanjing, China.}
		\thanks{Peng Wang (e-mail: peng.wang09@united-imaging.com), Guotao Quan (e-mail: guotao.quan@united-imaging.com) and Yanfeng Du (e-mail: yanfeng.du@united-imaging.com) are with CT RPA Department, United Imaging Healthcare Company Ltd., Shanghai, China.}
		\thanks{Dong Liang (e-mail: dong.liang@siat.ac.cn) and Yinsheng Li (e-mail: ys.li2@siat.ac.cn) are with Research Center for Medical Artificial Intelligence, and Key Laboratory of Biomedical Imaging Science and System, Shenzhen Institutes of Advanced Technology, Chinese Academy of Sciences, Shenzhen, China.}
		\vspace{-3em}
	}

\maketitle

\begin{abstract}  
	Dual-energy computed tomography (DECT) has been widely used to obtain quantitative elemental composition of imaged subjects for personalized and precise medical diagnosis. Compared with DECT leveraging advanced X-ray source and/or detector technologies, the use of the sequential-scanning data acquisition scheme to implement DECT may make a broader impact on clinical practice because this scheme requires no specialized hardware designs and can be directly implemented into conventional CT systems. However, since the concentration of iodinated contrast agent in the imaged subject varies over time, sequentially scanned data sets acquired at two tube potentials are temporally inconsistent. As existing material basis image reconstruction approaches assume that the data sets acquired at two tube potentials are temporally consistent, the violation of this assumption results in inaccurate quantification of material concentration. In this work, we developed sequenti\underline{a}l-s\underline{c}anning DE\underline{C}T imaging using high t\underline{e}mpora\underline{l} r\underline{e}solution image reconst\underline{r}uction and error-compensated m\underline{a}terial basis image genera\underline{tion}, ACCELERATION in short, to address the technical challenge induced by temporal inconsistency of sequentially scanned data sets and improve quantification accuracy of material concentration in sequential-scanning DECT. ACCELERATION has been validated and evaluated using numerical simulation data sets generated from clinical human subject exams and experimental human subject studies. Results demonstrated the improvement of quantification accuracy and image quality using ACCELERATION. 
\end{abstract}


\IEEEpeerreviewmaketitle

\section{Introduction}

Dual-energy computed tomography (DECT) has been widely used to obtain quantitative elemental composition of imaged subjects for personalized and precise medical diagnosis. Significant efforts have been made to implement DECT, such that, temporally consistent data sets at two different tube potentials can be acquired \cite{mccollough2020principles}. These DECT technologies can be broadly classified as two categories: advanced X-ray source-based technologies and detector-based technologies. X-ray source-based DECT equipped with two pairs of sources and detectors acquire data sets corresponding to two different spectrum at the same time \cite{flohr2006first}. Leveraging advanced source and detector technologies, tube potential can be switched from one to another at the speed in sub-millisecond level to acquire nearly consistent data sets \cite{zhang2011objective}. Two different beam filters are applied to split the X-ray beam into two halves along the longitudinal direction with different spectrum \cite{almeida2017dual,8080249}. Detector-based DECT equipped with energy-discriminating photon counting detectors acquire temporally consistent data sets with different spectrum \cite{symons2017photon}. Two different scintillation materials are manufactured along the depth direction of the detector to form the dual-layered detector, such that, two temporally consistent data sets, each with a different spectrum associated with the scintillation material can be readout \cite{lennartz2018dual}.

Although significant efforts have been made to implement DECT leveraging advanced X-ray source and/or detector technologies, the number of patients who can benefit from DECT may be limited partially because these high-end DECT may be limited to major clinical centers or research institutes in first-tier cities. To broaden the impact of DECT on clinical practice, DECT implemented with sequential-scanning data acquisition scheme may be a promising alternative because this scheme requires no specialized hardware designs and can be directly implemented into conventional CT systems. The feasibility of sequential-scanning DECT has been validated in static non-contrast CT imaging for the diagnosis of renal stones \cite{leng2015feasibility}.

\subsection{Technical Challenges in Sequential-Scanning Contrast-Enhanced DECT}

In sequential-scanning contrast-enhanced DECT, data sets corresponding to two different tube potentials are acquired separately. Since the concentration of iodinated contrast agent in the imaged subject varies over time, sequentially scanned data sets acquired at two tube potentials are temporally inconsistent. As existing material basis image reconstruction approaches for DECT assume that the data sets acquired at two tube potentials are temporally consistent, the violation of this assumption results in inaccurate quantification accuracy of material concentration and much degraded image quality. 



\subsection{Purpose and Innovation of This Work}

To address the technical challenge induced by temporal inconsistency of sequentially scanned data sets and improve quantification accuracy of material concentration in sequential-scanning contrast-enhanced DECT, in this work, we developed sequenti\underline{a}l-s\underline{c}anning contrast-enhanced DE\underline{C}T imaging using high t\underline{e}mpora\underline{l} r\underline{e}solution image reconst\underline{r}uction and error-compensated m\underline{a}terial basis image genera\underline{tion}, ACCELERATION in short. 

ACCELERATION first reconstructs several time-resolved images, each corresponding to an improved temporal resolution and lower temporal-averaging error, to resolve the temporal variation of the imaged subject using the short-scan data acquired at the first tube potential. Then, ACCELERATION resolves the temporal variation of the imaged subject using the short-scan data acquired at the second tube potential. The imaged subject corresponding to the selected reference time can be obtained via temporally extrapolating the time-resolved images corresponding to the first and second tube potentials.

To improve the quantitative accuracy and image quality of material basis images, material concentration can be quantified at the reference time using the pair of calculated temporally consistent images at two tube potentials and the developed error-compensated material basis image generation technique. The error-compensated material basis image generation models the degradation process into the forward model to train the generator using the data acquired for each individual subject. The degradation process includes the beam hardening effect due to polychromatic spectrum and structural error in the dual-energy images obtained from the developed high temporal resolution image reconstruction plus temporal extrapolation. Via properly modeling the degradation process, the forward model can maximally approximate the obtained dual-energy images, such that, the structural error in the obtained dual-energy images can be compensated for the improvement of accuracy in the obtained material basis images. 
	
The innovation in ACCELERATION is summarized as follows:
\begin{itemize}
	\item It achieves sequential-scanning dual-energy CT for contrast-enhanced imaging;
	\item It mitigates temporal inconsistency in terms of concentration of iodinated contrast agent;
	\item It explicitly models the degradation process into the forward model of the material basis image generation to properly compensate the structural error in the dual-energy images obtained from the developed high temporal resolution image reconstruction plus temporal extrapolation;
	\item It explicitly leverages the measured data of each individual subject to guarantee the accuracy of temporal matching and material basis image reconstruction for each individual subject.
\end{itemize}

In this work, we demonstrated that, a low-cost sequential-scanning data acquisition scheme to implement DECT has been achieved for contrast-enhanced imaging using the developed ACCELERATION technique. Results shown in this paper demonstrated the improvement of quantification accuracy and image quality using ACCELERATION. 

\section{ACCELERATION: Sequential-Scanning Dual-Energy CT Imaging Using Temporal Matching and Error-Compensated Material Basis Image Generation}

\subsection{Overall Framework}

Our goal is to address the technical challenge induced by temporal inconsistency of sequentially scanned data sets and improve quantification accuracy of material concentration in sequential-scanning DECT. The developed ACCELERATION includes two major techniques: (i) temporal matching using high temporal resolution image reconstruction and temporal extrapolation and (ii) material quantification using error-compensated material basis image generation. The temporal matching technique addresses the technical challenge induced by temporal inconsistency of sequentially scanned data sets in sequential-scanning DECT. The error-compensated material basis image generation improves material quantification accuracy and image quality of material basis images. 

\begin{figure*}
	\centering
	\includegraphics[width=0.9\textwidth, keepaspectratio=true]{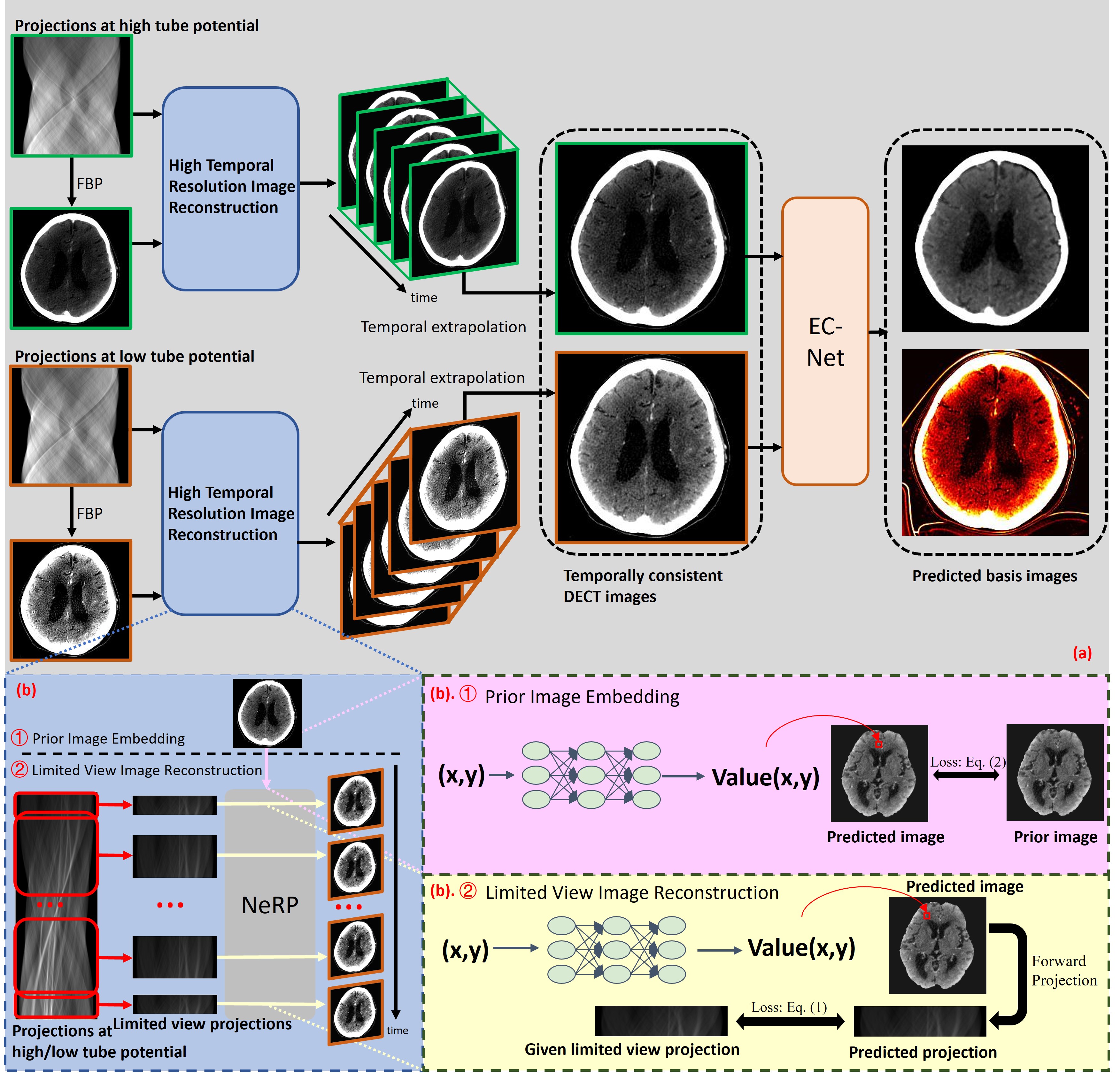}
	\caption{Overall framework of ACCELERATION; (a) The workflow of ACCELERATION; (b) The workflow of subject-specific high temporal resolution image reconstruction, \ding{172} and \ding{173} illustrate the details of the subject-specific high temporal resolution image reconstruction. "EC-Net" stands for the error-compensated material basis image generation network.}
	\label{fig:framework}
\end{figure*}

To make the subject temporally consistent, one has to first resolve the temporal variation of the subject using the data acquired at one of the tube potentials. To achieve this purpose, we developed a subject-specific deep time-resolved image representation to resolve the temporal variation of the subject during the time window to acquire the data at one of the tube potentials. The imaged subject corresponding to the selected reference time can be obtained via temporally extrapolating the time-resolved images corresponding to the data acquired at two tube potentials. To improve the quantitative accuracy and image quality of material basis images, material concentration can be quantified at the selected reference time using the pair of calculated temporally consistent images at two tube potentials and the developed error-compensated material basis image generation network. Fig. \ref{fig:framework} shows the overall framework of ACCELERATION. 

\subsection{High Temporal Resolution Image Reconstruction Using Data Acquired Over the Short-Scan Range}
\label{sec:recon}

For a system with a fixed gantry rotation speed, the only way to improve its temporal resolution is to use data acquired over a narrower angular range that violates the data sufficiency condition. For FBP, the data sufficiency condition specifies the minimal angular range as the short-scan angular range. Severe data-insufficiency-induced errors (a.k.a. limited view artifacts), can be observed using FBP from the data acquired over shorter than the short-scan range. Model-based iterative image reconstruction approaches leveraging prior image and sparsity \cite{tang2010temporal}, the low rankness of the spatial-temporal image matrix \cite{chen2015synchronized,li2018time} can be applied to reconstruct the entire imaged object without severe data-insufficiency-induced errors using a data sub-set acquired over up to one-fourth of the short-scan range. The four-dimensional digital subtraction angiography method reconstructs a three-dimensional image volume from the data acquired at a single projection view \cite{davis20134d} using both a pre-contrast scan (the one without contrast injection) and a post-contrast scan (the one with contrast injection). By incorporating the prior knowledge of the periodicity of the data-insufficiency-induced errors, eSMART-RECON further improves temporal resolution to 4-7.5 fps \cite{li2019enhanced}. To leverage the prior knowledge of the periodicity, eSMART-RECON requires that the acquired data set covers at least four scans (i.e. two forward scans plus two backward scans). By recasting the temporal resolution improvement task as a temporal extrapolation task, AIRPORT \cite{Li2023airport} improves the temporal resolution to 40 fps, the time window needed to acquire a single projection view data using a typical C-arm CBCT and the data acquired over the short-scan range.

CT image reconstruction using data acquired from a limited view range violates the data sufficiency condition and hence results in images with severe limited view artifacts unless appropriate regularization can be incorporated to constrain the otherwise ill-posed problem. However, it remains a technical challenge to appropriately design the regularization for better solving the limited view image reconstruction problem. Inappropriate regularization may lead to degraded quantitative accuracy (such as biased CT values) and/or image quality (degraded image sharpness, unnatural noise texture, image artifacts etc.). 

Instead of solving the CT value for each individual image voxel, one can also represent the image-to-be-reconstructed as a deep convolutional neural network \cite{ulyanov2018deep,barbano2022educated,shu2022sparse} or a deep fully connected network \cite{sitzmann2020implicit,molaei2023implicit,Reed_2021_ICCV,shen2022nerp,zhang2023dynamic,song2023piner,lee2024iterative} with learnable parameters to regularize the image reconstruction problem using the architecture of the deep neural network. The network parameters can then be determined by best fitting the forward model of the image estimation to the measured data corresponding to a limited view range. After training the network using the data corresponding to the imaged subject, the deep neural networks with fully connected layers learns a smooth mapping from the image voxel coordinates domain to the corresponding intensity value domain. As the smoothness has been implicitly enforced from one image voxel to other nearby voxels, CT image noise and artifacts can be effectively mitigated in the reconstructed image represented by the learned smooth mapping, such that, CT image reconstruction problems under nonideal data acquisition conditions can be effectively solved. The high temporal resolution image reconstruction using deep neural networks with fully connected layers includes two steps: prior image embedding (see in Fig.~\ref{fig:framework} (b). \ding{172}) and limited view image reconstruction (see in Fig.~\ref{fig:framework} (b). \ding{173}).  
 
\subsubsection{Prior Image Embedding}

In the prior image embedding step, a deep neural network with multiple fully connected layers $\mathcal{M}_{\mathbf{\Theta}}$ is trained to express the mapping from image voxel coordinates to intensity values associated with the image voxel, that is, $\mathcal{M}_{\mathbf{\Theta}}(c_j) \rightarrow v_j$, where $j$ denotes the coordinate index in the image spatial domain. The optimization problem in the prior image embedding step can be formulated as follows:
\begin{align}
	\label{eq:prior-embedding}
	{\mathbf{\Theta}}^{*}_0=\arg\min_{\mathbf{\Theta}} \frac{1}{N}\sum_{j=1}^{N} \left(\mathcal{M}_{\mathbf{\Theta}}(c_j) - v_j\right) ^{2},
\end{align}
where $N$ denotes the total number of voxels in the CT image. The anatomical structure of the temporally averaging image over the short-scan range (a.k.a. prior image) can be implicitly embedded into the network parameters which can be considered as a warm initialization to represent the imaged subject associated to a limited view range in the next limited view image reconstruction step. 

\subsubsection{Limited View Image Reconstruction}

The optimization problem in the limited view image reconstruction step can be formulated as follows:
\begin{align}
	\label{eq:limited-view-reconstruction}
	{\mathbf{\Theta}_t}^{*}&=\arg\min_{\mathbf{\Theta}_t} \frac{1}{|S_t|}\sum_{i\in S_t} \left( \sum_{j=1}^N a_{ij} \mathcal{M}_{\mathbf{\Theta}_t}(c_j) - s_i \right)^2,
\end{align}
where $\mathbf{\Theta}_t$ denotes the network parameters corresponding to the $t$-th time frame, namely a sub-set of measured data corresponding to a limited view range, $t = 1,2,\cdots,T$. $S_t$ denotes the set of measurement corresponding to the $t$-th time frame. The detailed implementation of $\{S_t\}$ is specified in Sec. \ref{HTRdetails}. $|\cdot|$ denotes the number of elements in the set, and $T$ denotes the number of reconstructed time frames which is a hyper-parameter to be optimized for the best reconstruction accuracy. Note that, the learned network parameters $\mathbf{\Theta}^{*}_0$ in the prior image embedding (Eq.~(\ref{eq:prior-embedding})) are used as the initialization in solving these optimization problems defined in Eq. (\ref{eq:limited-view-reconstruction}) for each time frame.

\subsection{Temporal Matching by Extrapolation}
\label{sec:extra}

Suppose that scanning at the first tube potential is performed during $[0,T_{2}]$, the tube potential is changed during $[T_{2},T_{4}]$, and scanning at the second tube potential is performed during $[T_{4},T_{6}]$. The variation of iodine concentration of the imaged subject underwent intravenous contrast injection at each image voxel can be approximately considered as general Gamma-Variate functions \cite{preslar1991new} over a relatively long duration (normally longer than 30 seconds). Given the short duration of data acquisition (e.g. about 0.2 seconds to acquire a short-scan data set using a state-of-the-art multi-detector-row CT scanners), the variation of iodine concentration and CT attenuation values within such short duration can be legitimately assumed as linear variation.

Based on the $T$ time-resolved CT images obtained by solving Eq.~(\ref{eq:limited-view-reconstruction}), for each voxel in the image, a linear least squares fitting is used to find a linear representation of the temporal variation of intensity value at the given image voxel. CT images corresponding to $T_3$ (the midpoint of $[T_{2},T_{4}]$) and the first tube potential can then be determined by linear extrapolation of each linear representation of the temporal variation of intensity value at the each image voxel corresponding to time-resolved images during $[0,T_{2}]$. Similarly, CT images corresponding to $T_3$ and the second tube potential can then be determined by linear extrapolation of each linear representation of the temporal variation of intensity value at the each image voxel corresponding to time-resolved images during $[T_{4},T_{6}]$. Then, temporal inconsistency in CT images corresponding to both the first and the second tube potentials are maximally mitigated at $T_3$. Material quantification can then be performed to obtain quantitative material concentration using the above extrapolated dual-energy images at $T_3$.

\subsection{Material Quantification Using Error-Compensated Material Basis Image Generation}

Through the developed high temporal resolution image reconstruction and temporal extrapolation, we can obtain CT images at two tube potentials with the nearly consistent iodine concentration, which can be used for material quantification. Due to the instability of material quantification, structural errors in the reconstructed dual-energy CT images obtained from high temporal resolution image reconstruction plus temporal extrapolation developed in Sec.~\ref{sec:recon} and Sec.~\ref{sec:extra} will be greatly amplified \cite{alvarez1976energy}, resulting in erroneous material basis images. 

To mitigate the errors in the material basis image, in this paper, instead of applying conventional image-domain material quantification procedure to obtain material basis images, an error-compensated material basis image generation network, EC-Net, has been designed and properly trained to improve quantitative accuracy and image quality of material basis images for each individual subject. The error-compensated material basis image generation models the degradation process into the forward model to train the generator using the data acquired for each individual subject. The degradation process includes the beam hardening effect due to polychromatic spectrum and structural error in the dual-energy images obtained from the developed high temporal resolution image reconstruction plus temporal extrapolation. Via properly modeling the degradation process, the forward model can maximally approximate the obtained dual-energy images, such that, the structural error in the obtained dual-energy images can be properly compensated to further improve the accuracy of the obtained material basis images.

EC-Net is trained to learn the subject-specific material quantification mapping including two sub-mappings $\mathcal{G} \circ \mathcal{D}$. Here $\mathcal{G}$ denotes the material basis image generation mapping from randomly sampled noise image to ideal material basis images. $\mathcal{D}$ denotes the dual-energy CT image degradation mapping from ideal material basis images to dual-energy CT images from the developed high temporal resolution image reconstruction plus temporal extrapolation. The material basis image generation mapping $\mathcal{G}_\mathbf{\Phi}$ is learned for each subject individually to maximally avoid the generalizability error induced by a model trained using a relative large data cohort. The dual-energy CT image degradation mapping $\mathcal{D}_\mathbf{\Psi}$, on the other hand, can be trained using a relative large data cohort to represent the features of structural errors represented in a large group of dual-energy CT images. To implement this idea, model parameters $\mathbf{\Psi}$ are determined by solving the following problem using a relatively large training data set and will be frozen during the training of the material basis image generation mapping $\mathcal{G}_\mathbf{\Phi}$. 

The optimization problem for determining $\mathbf{\Psi}$ can be formulated as follows:
\begin{align}
	\label{eq:frozenLayer}
	\mathbf{\Psi}^{*} &= \arg\min_{\mathbf{\Psi}} \| \mathcal{D}_\mathbf{\Psi}(\hat{\mathbf{X}}) - \tilde{\mathbf{X}} \|_2^{2}, \\ \nonumber
	\hat{\mathbf{X}} &= [\hat{\mathbf{x}}_L, \hat{\mathbf{x}}_H], ~~ \tilde{\mathbf{X}} = [\tilde{\mathbf{x}}_L, \tilde{\mathbf{x}}_H],
\end{align}
where $\hat{\mathbf{x}}_L$ and $\hat{\mathbf{x}}_H$ denote the simulated dual-energy CT images using ideal material basis images at the low and high tube potentials respectively, $\tilde{\mathbf{x}}_L$ and $\tilde{\mathbf{x}}_H$ denote the reconstructed dual-energy CT images using the high temporal resolution image reconstruction and temporal extrapolation developed in Sec.~\ref{sec:recon} and Sec.~\ref{sec:extra} respectively.

The optimization problem for determining $\mathbf{\Phi}$ can be formulated as follows:
\begin{align}
	\label{eq:trainLayer}
	\mathbf{\Phi}^{*}&= \arg\min_{\mathbf{\Phi}} \| \mathcal{D}_{\mathbf{\Psi}^{*}} \circ \mathcal{S} \circ \mathcal{G}_\mathbf{\Phi}(\mathbf{n}) - \tilde{\mathbf{X}} \|_2^{2}, \\ \nonumber
\end{align}
where $\mathbf{n}$ denotes the random values sampled from the uniform distribution. Once the material basis image generation mapping has been trained for each individual subject, the optimal material basis images can be obtained via $\mathcal{G}_{\mathbf{\Phi}^*}(\mathbf{n})$.

$\mathcal{S}$ represents the operator for synthesizing DECT images showing as follows:
\begin{align}
	\label{eq:synthesis}
	\hat{\mathbf{X}} &= \mathcal{S}\left(\mathcal{G}_\mathbf{\Phi}(\mathbf{n}) \right), \\ \nonumber
	\hat{\mathbf{X}}_{j,k^{'}} &:= \sum_i \mathbf{a}^{+}_{i,j} \hat{\mathbf{Y}}_{i,k^{'}}, \\ \nonumber
	\hat{\mathbf{Y}}_{i,k^{'}} &:= -\log\left( \sum_e \Omega_{k^{'},e} \exp\left( -\sum_k \hat{\mathbf{P}}_{i,k} b_{k,e}\right) \right), \\ \nonumber
	\hat{\mathbf{P}}_{i,k} &:= \sum_j \mathbf{a}_{i,j} \left[\mathcal{G}_\mathbf{\Phi}(\mathbf{n})\right]_{j,k},
\end{align}
where $\hat{\mathbf{Y}}$ denotes the modeled dual-energy post-log sinogram data, $\Omega_{k^{'},e}$ denotes the normalized photon number corresponding to $k^{'}$-th tube potential at the $e$-th energy index, $b_{k,e}$ denotes the mass attenuation coefficient of $k$-th basis material at $e$-th energy index, $\mathbf{a}_{i,j}$ denotes $i$-, $j$-th element of the system matrix $\mathbf{A}$, $\mathbf{a}^{+}_{i,j}$ denotes $i$-, $j$-th element of $\mathbf{A}^{+}$ which represents the pseudo inverse of $\mathbf{A}$. Here we used the standard FBP algorithm to implement $\mathbf{A}^{+}$.

\subsection{Implementation Details}

All our experimental implementations are based on PyTorch, and the differentiable forward projection operators in neural network are based on LEAP in \cite{kim2023differentiable}. All computations were performed on a workstation equipped with an Intel(R) Xeon(R) Platinum 8160 CPU @ 2.10GHz and an RTX 3090 GPU with 24GB of memory.

\subsubsection{Data Pre-processing}
\label{prep}

To decouple the interference between static bony structures (such as skull) and temporal varying contrast-enhanced structures (such as iodinated cerebral tissues or vessels) in the sinogram domain, the contribution of static bony structures in the sinogram domain has been subtracted from the measured data prior to the following high temporal resolution image reconstruction. To achieve this purpose, we first applied FBP to reconstruct the measured data acquired at two tube potentials. We then applied a thresholding-based segmentation approach to discriminate bony structures from other tissues. After forward projecting the segmented bony structures, we subtracted the contribution of static bony structures in the sinogram domain from the measured data. 

\subsubsection{High Temporal Resolution Image Reconstruction}
\label{HTRdetails}

The input of the high temporal resolution image reconstruction is the short-scan range data at one of the tube potentials acquired during $[0,T_{2}]$ or $[T_{4},T_{6}]$, and its output corresponds to $T$ time frames at equally spaced intervals within the time intervals \([0, T_2]\) or \([T_4, T_6]\). In this phase, the $\mathcal{M}_{\mathbf{\Theta}}$ consists of a Fourier feature embedding layer \cite{shen2022nerp} and an eight fully-connected layers. The width of fully-connected layers is 256 and the activation function is sinusoidal activation functions \cite{sitzmann2020implicit}. For prior image embedding, we used Adam optimizer with an initial learning rate of $1\times10^{-4}$, and the maximum number of training iterations is set to 1500. For limited view image reconstruction, the initial learning rate of Adam optimizer is set to $5\times10^{-6}$ for the first and last time frames, and $5\times10^{-9}$ for the middle time frames, linearly decreasing from peripheral to middle frames. The maximum number of training iterations is set to 2000.

During limited view image reconstruction, elements in the set of $\{S_t\}$ are determined flexibly, allowing for an overlapped data usage in the reconstruction of different time frames. For any specific value of $T$, total number of $T$ central views can be identified. The set of projection in the reconstruction of $t$-th frame was determined by maximally utilizing all available data equally distributed around the central view of $t$-th frame. For the time frames close to the boundaries of the total angular range, fewer data were used, while for those in the middle, more data were used. Obviously, this scheme can reduce limited view artifacts in middle time frames but introduces temporal-averaging errors \cite{chen2015synchronized}. 

$\{S_t\}$ is empirically designed to solve the high temporal resolution image reconstruction problem. The specific implementation of $\{S_t\}$ is not necessary to be generalizable to solve other image reconstruction problems in other imaging tasks, therefore, it should be optimized for each individual image reconstruction problem or imaging task. In this work, $T=5$ was empirically selected. 


\begin{figure*}
	\centering
	\includegraphics[width=0.8\textwidth, keepaspectratio=true]{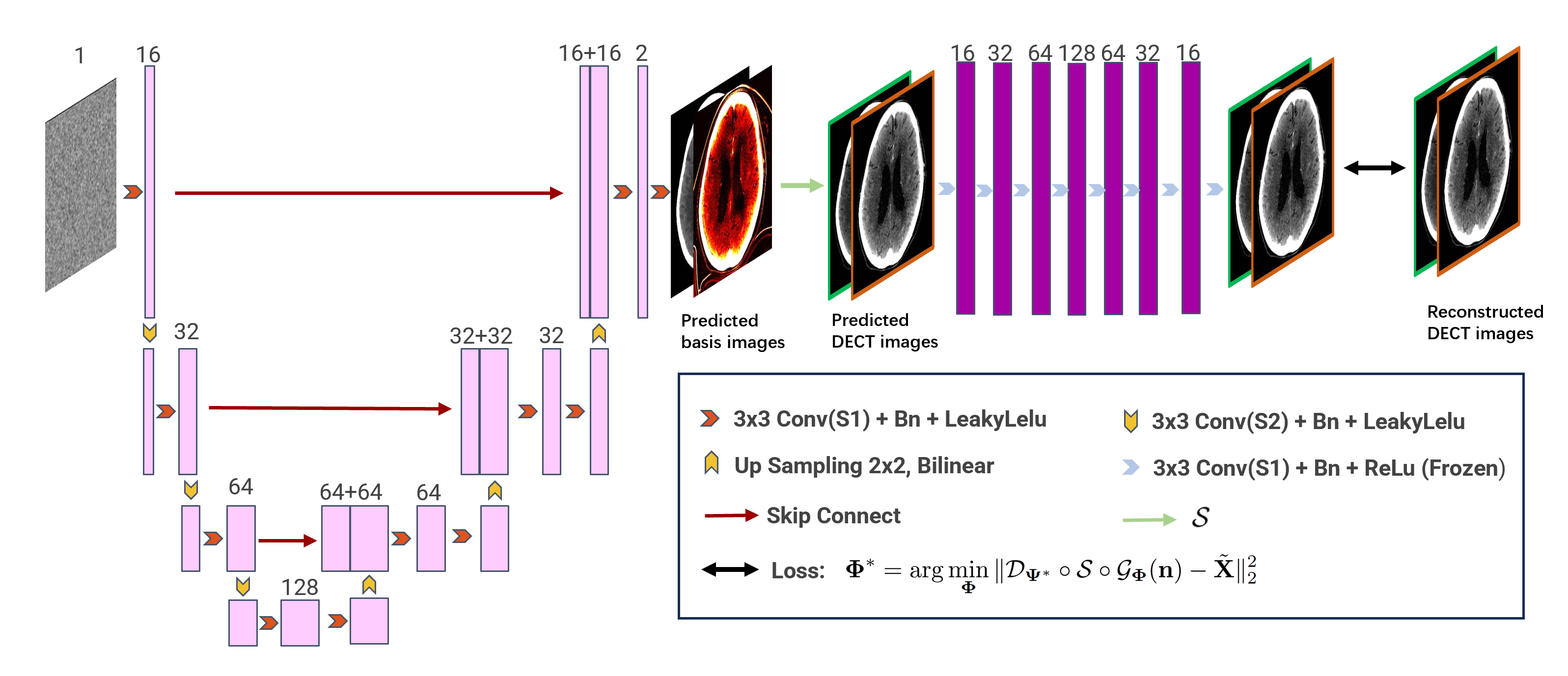}
	\caption{The detailed structure of EC-Net. The temporally consistent reconstructed DECT images as input. The predicted basis images as output.}
	\label{fig:ec-net}
\end{figure*}

\subsubsection{Temporal Extrapolation}

Based on the reconstructed time-resolved CT images obtained from the high temporal resolution image reconstruction, temporal variation of each pixel's attenuation values is linearly fitted using the least square method. The value at \( T_3 \) is calculated from the fitted line, representing the temporally extrapolated attenuation value at \( T_3 \). After temporal extrapolation, temporally consistent CT images under different tube potentials were obtained.

\subsubsection{EC-Net}

The EC-Net consists of an encoder-decoder convolutional neural network $\mathcal{G}$ (the pink section in Fig.~\ref{fig:ec-net}) and a traditional seven-layer convolutional neural network $\mathcal{D}$ (the purple section in Fig.~\ref{fig:ec-net}). The model parameters of $\mathcal{D}$ is pre-trained using 3000 samples and the model parameters of $\mathcal{D}$ are frozen in solving Eq.~(\ref{eq:trainLayer}).

For $\mathcal{G}$, the network structure is shown in Fig.~\ref{fig:ec-net} and hyperparameters follow the image denoising settings in \cite{ulyanov2018deep}. The maximum number of training iterations is set to 1500, and the learning rate of Adam optimizer is set to $2\times10^{-3}$. For $\mathcal{D}$, the number of convolutional kernels per layer is specified in the figure. Each layer includes a batch normalization operation and uses ReLU as the activation function. 

\section{Materials and Methods to Validate and Evaluate ACCELERATION}

To validate the proposed ACCELERATION, we employed numerical simulations with known ground truth to assess its performance and accuracy. Additionally, data acquired from human subjects were introduced to verify the clinical feasibility of ACCELERATION. ACCELERATION was compared against several baseline methods.

\subsection{Numerical Simulation Studies with Known Ground Truth}

\begin{figure*}
	\centering
	\includegraphics[width=0.9\textwidth, keepaspectratio=true]{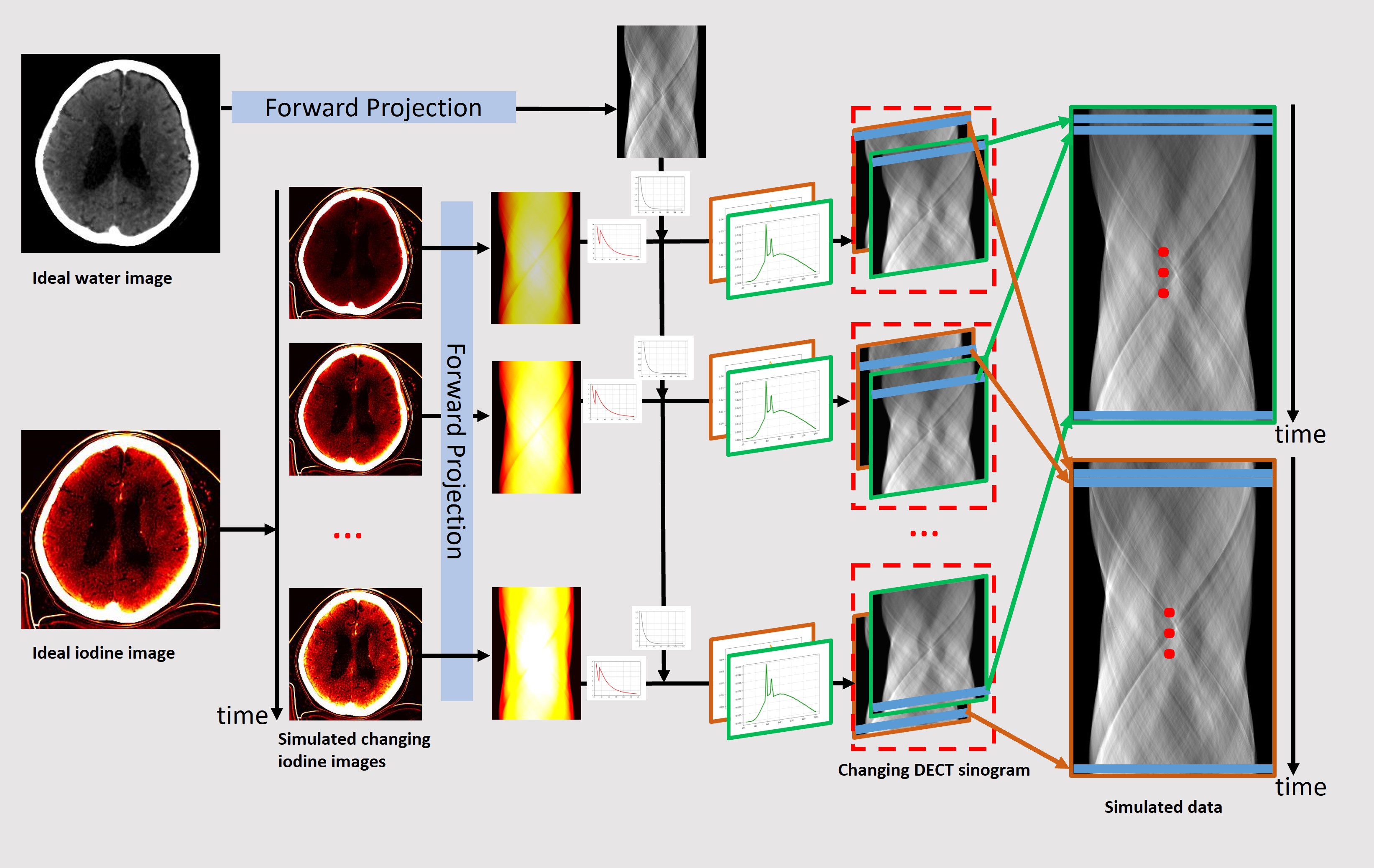}
	\caption{The workflow of data generation in numerical simulation studies.}
	\label{fig:simulation}
\end{figure*}


\subsubsection{Data Collection}

To obtain numerical simulation data with known ground truth, a pair of cerebral water and iodine basis images was used as the ground true material basis images for the simulation experiments. The ground true material basis images were obtained using a clinical high-end DECT and a clinical contrast-enhanced cerebral CT angiography data acquisition protocol. The 512$\times$512 CT images were simulated using a fan-beam CT scanner with 384 detectors, 1000 views, 210 degrees for a short-scan range, and data acquisition time of 1 second. 

\subsubsection{Training and Validation Data Generation}

To simulate the variation of iodine contrast agent in the brain of the subject, we set that the concentration of the iodine contrast agent in blood vessels linearly increased to 2.5 times the basis image in the interval \([0, T_6]\), while the soft tissue increased to 1.5 times the basis image. In CT images at 140 \text{kV}, vascular changes are approximately 80 \text{HU}, while soft tissue changes are about 20-30 \text{HU}.

The details of data generation in numerical simulation studies are shown in Fig.~\ref{fig:simulation}.
Based on the simulated variation of iodine concentration, we obtained iodine basis images corresponding to each view angle during the short scans \([0, T_2]\) and \([T_4, T_6]\), and performed forward projection for iodine basis images corresponding to each view angle using the forward projection operator in \cite{kim2023differentiable}. We performed forward projection for water basis images and assumed that the water basis image remains unchanged. Note that it is a legitimate assumption because in contrast-enhanced CT imaging tasks, the variation of iodine concentration takes solo responsibility for the variation of attenuation values. Based on the energy-dependent attenuation coefficients of basis materials and simulated X-ray spectrum, line integrals corresponding to each view angle under 80/140 \text{kV} were simulated. By performing filtered back projection (FBP) on each short-scan line integral data, the ground truth CT images at each time point (corresponding to a short-scan angule range) under 80/140 \text{kV} can be obtained. The sequential-scanning DECT data in \([0, T_2]\) and \([T_4, T_6]\) can be constructed by extracting the line integral data corresponding to each view angle and combining them into two short-scan sinogram data under 80/140 \text{kV}.

We used 150 samples to generate the training data to learn the dual-energy CT image degradation mapping $\mathcal{D}_\mathbf{\Psi}$. For each sample, 20 values were randomly sampled from a uniform distribution over \([0, 200\%]\). These sampled values represented the variation in iodine concentration within the interval \([0, T_6]\). We generated the corresponding training data for high temporal resolution reconstruction. After high temporal resolution image reconstruction and temporal extrapolation, reconstructed CT images at \( T_3 \) with structural errors are paired with the ground truth, and were used for training the degradation mapping $\mathcal{G}_\mathbf{\Phi}$.

\subsection{Experimental Human Subject Studies}

In addition to numerical simulations studies, in vivo human subject studies have been conducted to further demonstrate the technical feasibility of ACCELERATION. Data sets of one human subject (male, 44 years old) acquired with an institutional review board approval and written informed consent at Shanghai Jiading Central Hospital were retrospectively analyzed. The data were acquired using a MDCT system (uCT 960+, United Imaging Healthcare, Shanghai, China). Both the data acquisition and contrast agent injection protocols were approved by the institution. Projection data were processed with the needed data correction using the proprietary software toolkit provided by the vendor.

\subsection{Baseline Methods for Performance Comparison} 

In the high temporal resolution reconstruction phase, we selected FBP, SIRT \cite{trampert1990simultaneous}, PICCS \cite{tang2010temporal}, and SMART \cite{chen2015synchronized} as baseline methods. We provided the SIRT, PICCS, and SMART methods with the same prior image as the initial images for iteration as used in the proposed method, and the same projection selection method and CT image reconstruction frame count were used for fair comparison. We optimized the hyperparameters of the baseline methods as much as possible under our imaging conditions. In PICCS, \(\lambda_1\) was set to 0.91, and \(\lambda_2\) was set to 0.09. In SMART, singular values greater than 5\% of the maximum singular value were retained. Other hyperparameters for each method were set according to the recommendations in the reference articles. In the material quantification phase, we introduced the direct inversion method \cite{alvarez1976energy} after FBP reconstruction of DECT images and direct inversion with the proposed high temporal resolution reconstruction and temporal extrapolation to demonstrate the effectiveness of the proposed workflow. 

\subsection{Metrics for Quantitative Performance Assessment}

The normalized root mean square error (nrMSE) was used to quantitatively assess the similarity between the reconstructed image and the ground truth.
\[
\text{nrMSE} = \frac{\sqrt{\frac{1}{N} \sum_{i=1}^{N} (x_i - y_i)^2}}{\sqrt{\frac{1}{N} \sum_{i=1}^{N} y_i^2}} \times 100\%,
\]
where \( x_i \) is the pixel value of the reconstructed image, \( y_i \) is the pixel value of the ground truth image, and \( N \) is the total number of pixels, and the denominator is the root mean square value of the ground truth image. Additionally, we subtracted 1000 \text{mg/ml} from the water basis image to amplify the nrMSE, thereby more effectively reflecting the error between the reconstructed water basis image and the ground truth.


\section{Results}

\subsection{Numerical Simulation Studies}

\subsubsection{Validation of Numerical Optimization}

\begin{figure}[h]
	\centering
	\includegraphics[width=0.45\textwidth, keepaspectratio=true]{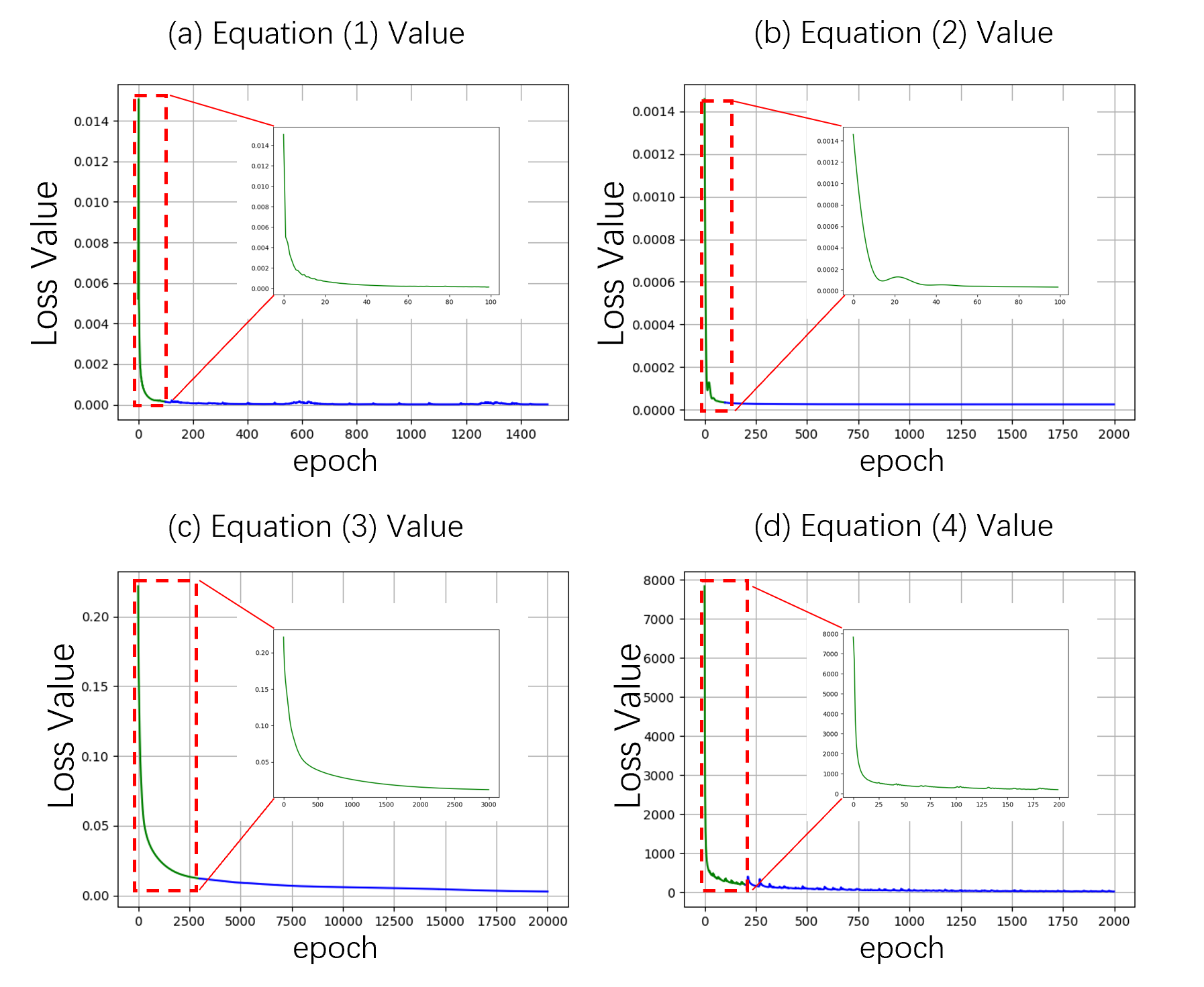}
	\caption{The change of loss function value in Eq.~(\ref{eq:prior-embedding}) - Eq.~(\ref{eq:trainLayer}) with respect to each epoch.}
	\label{fig:loss}
\end{figure}

Empirical convergence of the optimization process of ACCELERATION is shown in the change of loss function value of the corresponding prediction with respect to each epoch in Fig.~\ref{fig:loss}. As shown in Fig.~\ref{fig:loss}, the loss function value quasi-monotonically decreases during the optimization. The plateau of the total loss values indicates the empirical convergence of the numerical optimization process.

\subsubsection{Assessment of High Temporal Resolution Image Reconstruction}

\begin{figure}[h]
	\centering
	\includegraphics[width=0.45\textwidth, keepaspectratio=true]{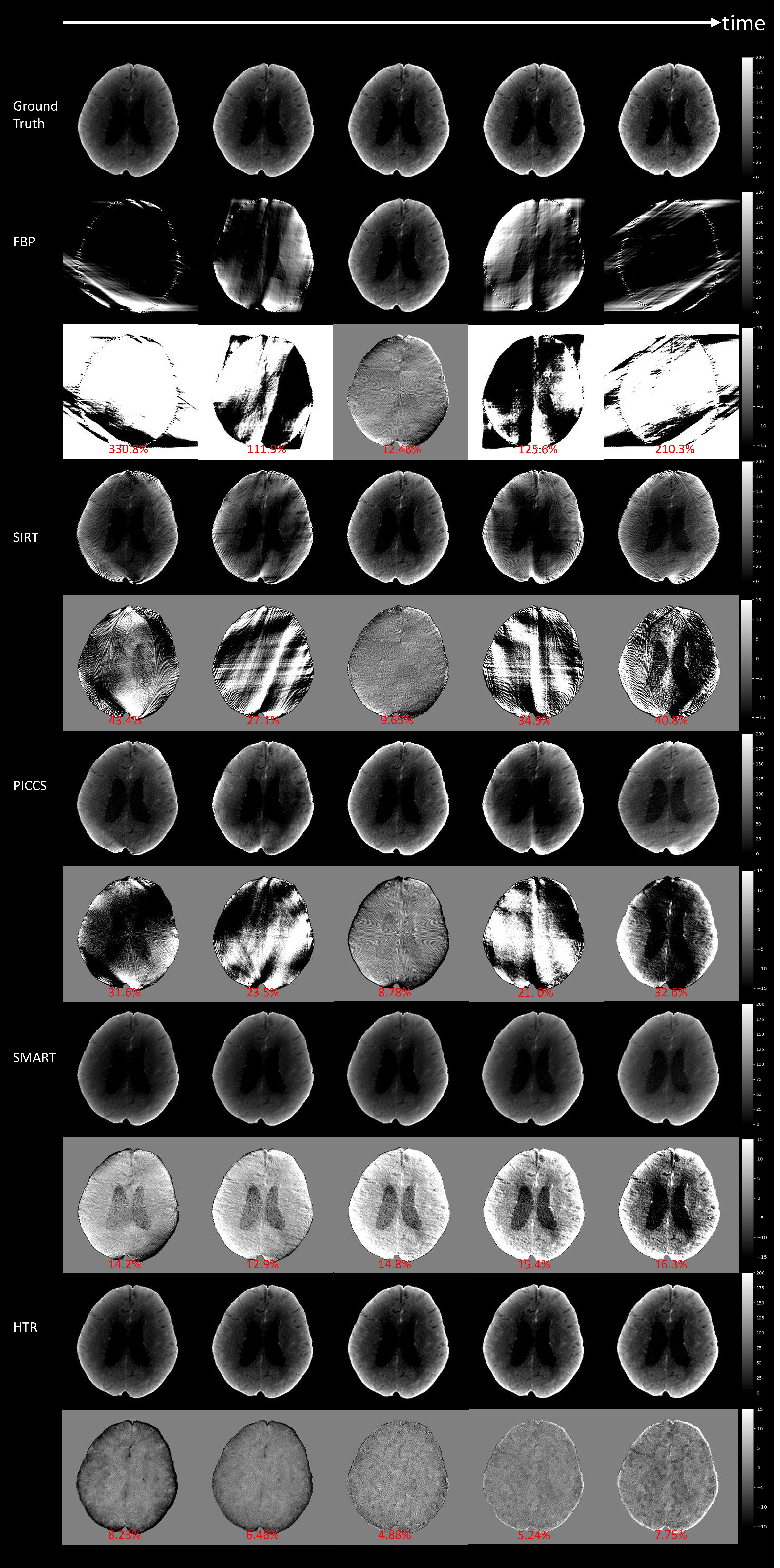}
	\caption{Numerical phantom images generated from Ground Truth, FBP, SIRT, PICCS, SMART and the proposed HTR. Images and difference images are displayed with W/L: 200/100 HU and 30/0 HU respectively. The nrRMSE for each of these images was calculated.}
	\label{fig:htr}
\end{figure}

Projections at 80/140 kV generated from numerical simulation studies were used to evaluate the algorithms in the high temporal resolution image reconstruction. The ground true images at different time points are shown in the first row of Fig.~\ref{fig:htr}. The baseline methods, SIRT, PICCS, and SMART, are shown in the 2nd, 3rd, and 4th rows, respectively. Clearly, FBP exhibits severe artifacts for limited view reconstruction. By using prior images as the initial image, SIRT shows some alleviation of artifacts compared with FBP but still presents noticeable limited view artifacts. PICCS algorithm introduces compressed sensing based on sparse-promoting regularization, significantly reducing limited view artifacts in most time frames. However, it exhibits notable errors as shown in the difference images and fails to clearly depict the temporal variation of the imaged object. SMART algorithm leverages low rank regularization on the spatial-temporal image matrix, nearly eliminating limited view artifacts and qualitatively depict the temporal variation of the imaged object. However, it underestimates attenuation values, resulting in quantitative errors that will be further amplified during temporal extrapolation and material quantification. The proposed high temporal resolution image reconstruction, HRT in short, resolves the temporal variation of the imaged object with sufficient quantitative accuracy and image quality.


To evaluate the reconstruction accuracy statistically, we performed 200 rounds of data generation and high temporal resolution image reconstruction under the condition of an 80\% change in iodine concentration. For the blood vessel region, the absolute error is 7.6 [0, 15.9] HU (mean value and 95\% confidence interval respectively). For the cerebral tissue region, the absolute error is 2.4 [0, 6.2] HU.

\begin{figure}[ht]
	\centering
	\includegraphics[width=0.5\textwidth, keepaspectratio=true]{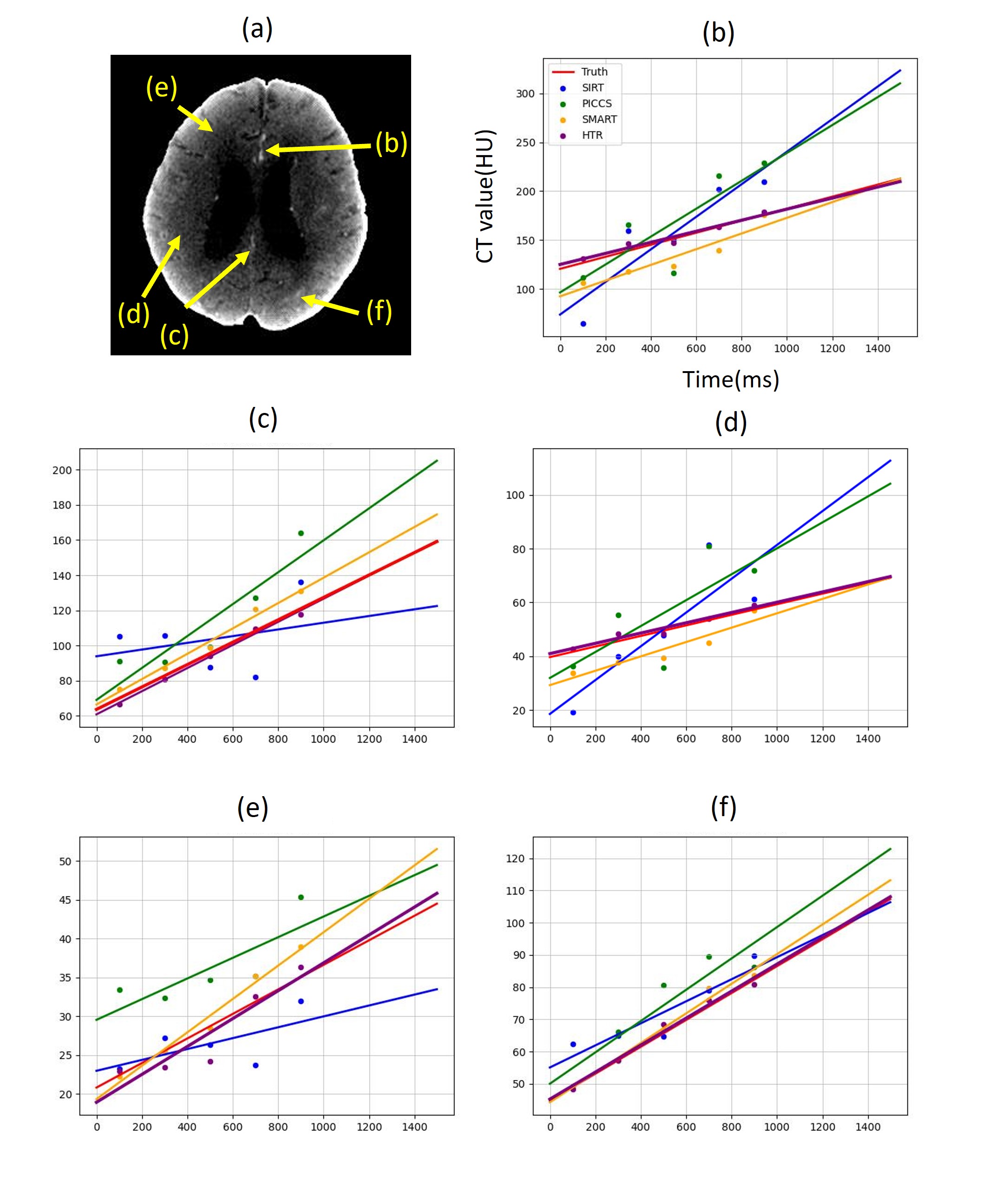}
	\caption{The reconstructed CT values and the accuracy of linear temporal extrapolation for different methods in various ROI are shown. (a) displays the locations of the selected ROIs presented in (b), (c), (d), (e), and (f). The horizontal axis represents time. HRT is performed during [0,1000] ms and HRT images is linearly extrapolated to 1500 ms for temporal matching.}
	\label{fig:htr-curves}
\end{figure}

Fig. \ref{fig:htr-curves} shows the comparison of the accuracy of high temporal image resolution reconstruction and linear extrapolation for baseline methods and the proposed method. Results shown in the figure demonstrate that the proposed method achieves high reconstruction accuracy in both vascular regions (subfig. b, c) and cerebral tissue regions (subfig. d, e, f).

\subsubsection{Performance Dependence of High Temporal Resolution Image Reconstruction on the Variation Rate of Contrast Concentration}

\begin{figure}[ht]
	\centering
	\includegraphics[width=0.5\textwidth, keepaspectratio=true]{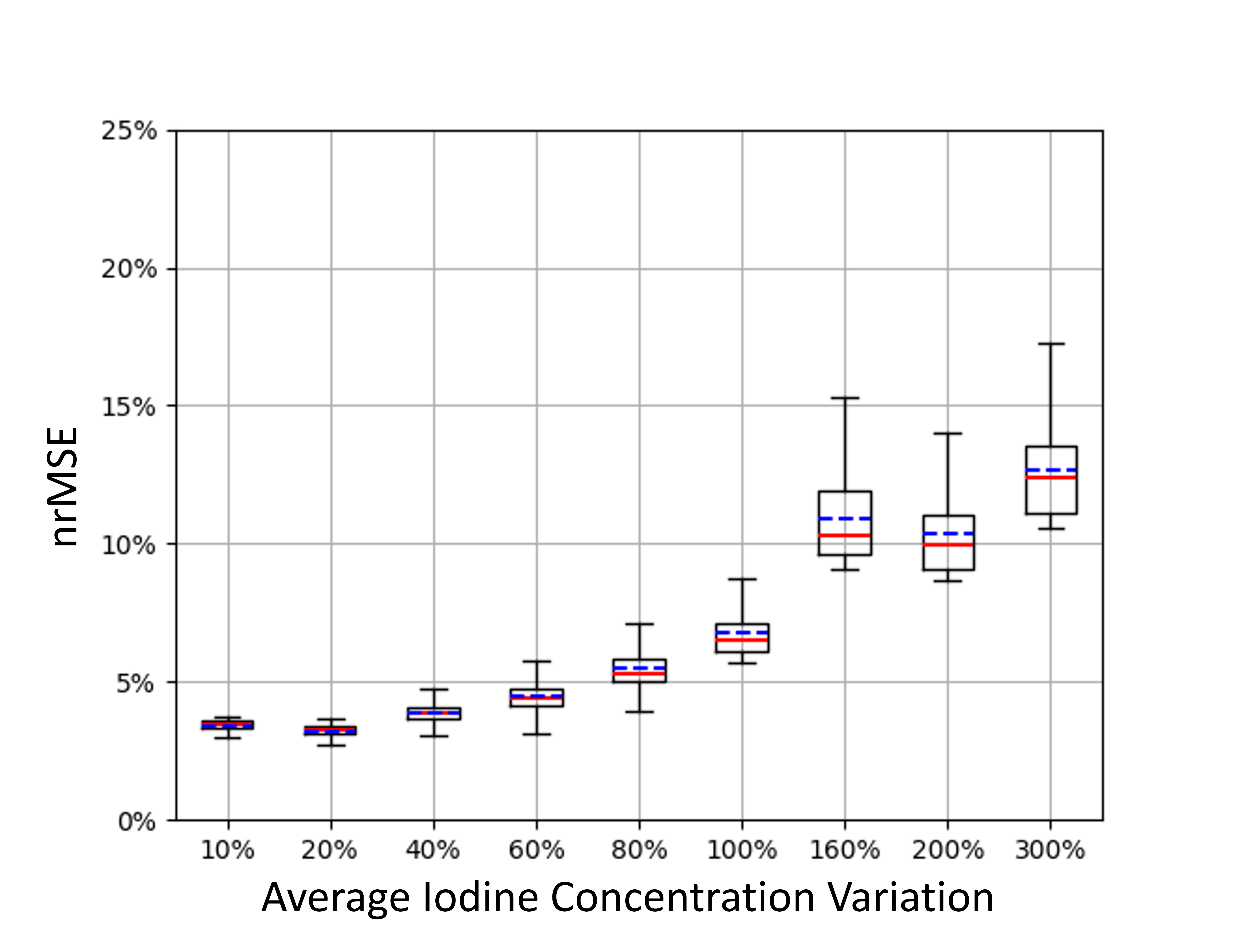}
	\caption{The performance dependence of HTR on the variation rate of contrast agent concentration.}
	\label{fig:dependence}
\end{figure}

The variation rate of contrast agent concentration within the brain affects the accuracy and feasibility of the proposed method. We investigated how the accuracy of our method changes with respect to different variation rates of contrast agent concentration. Fig.~\ref{fig:dependence} shows the accuracy (quantified by nrMSE) of our method when the average iodine concentration changes by 10\% - 300\% during the scan duration of 1000 ms. It can be observed that when the variation rate of contrast agent concentration is less than 60\%, the quantitative error of our method remains within 5\%. For variation rates between 60\% and 150\%, the error remains within 10\%. Even with very drastic changes in contrast agent concentration (e.g. peak CT values caused by the injection of contrast agent in proximal arteries), the error remains within 15\%. 

\subsubsection{Assessment of Material Quantification}

\begin{figure}[h]
	\centering
	\includegraphics[width=0.45\textwidth, keepaspectratio=true]{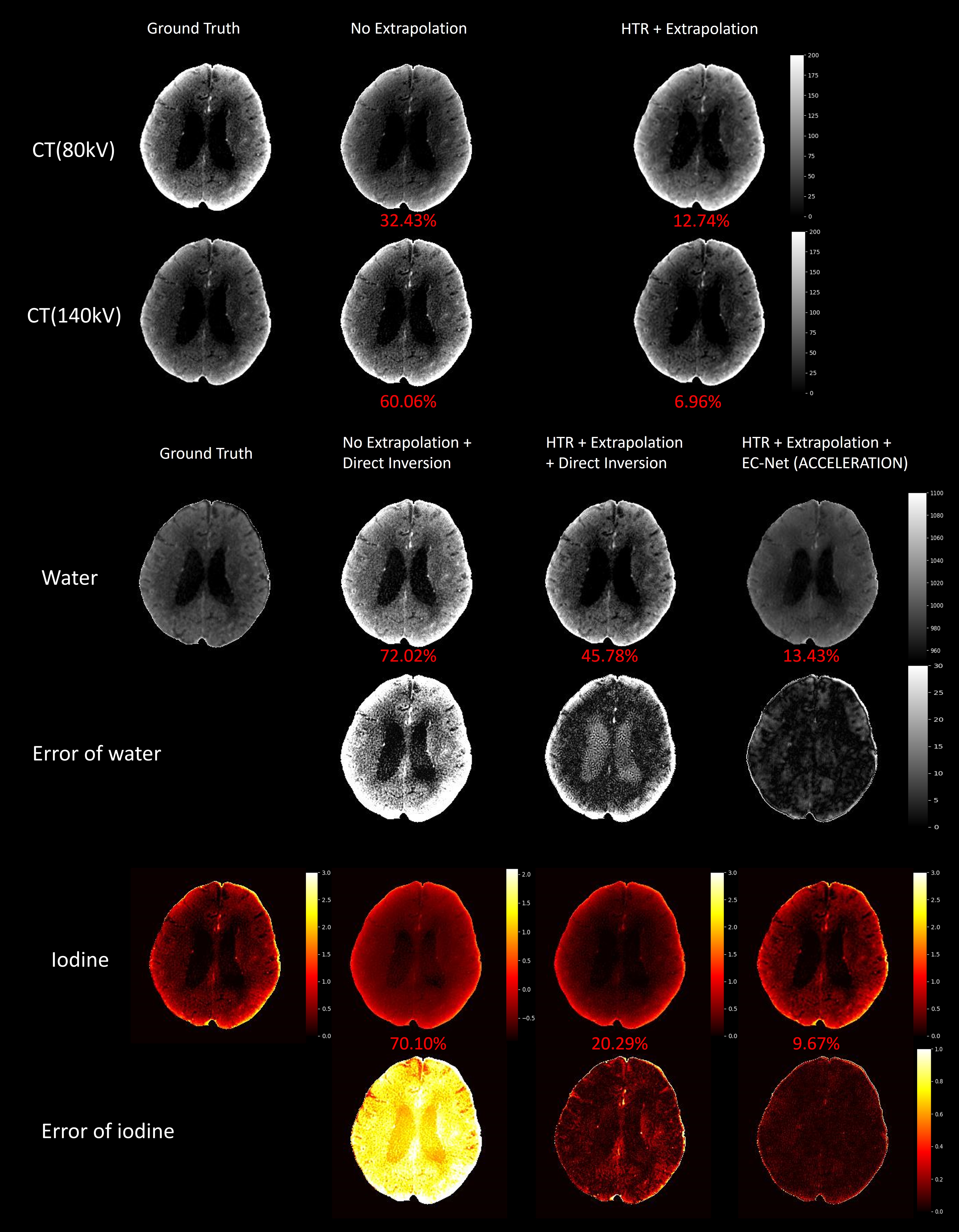}
	\caption{Material quantification results using numerical simulation studies. The first and second rows display CT images at 80 kV and 140 kV, respectively. The third and fourth rows present material basis images in mg/ml, obtained from the images in the first two rows. The first column shows the ground truth, the second column shows the material basis images without temporal extrapolation. The third column shows the results using the proposed HTR and direct inversion. The fourth column shows the results using ACCELERATION. nrMSE for each of the material basis images was calculated.}
	\label{fig:md}
\end{figure}

Material basis images obtained in the numerical simulation studies have been shown in Fig. \ref{fig:md}. In the second column, without using the proposed HTR and temporal extrapolation, material basis images obtained by the direct image-domain inversion are completely incorrect. In the third column, material basis images obtained by HTR, temporal extrapolation, and direct image-domain inversion, demonstrated amplified noise level, biased quantification of material concentration, and substantial loss of details partially due to the illness of the direct image-domain inversion. In the fourth column, material basis images obtained by HTR, temporal extrapolation and EC-Net, material basis images are reconstructed with sufficient quantitative accuracy and image quality. 
Note that we did not present material basis images obtained from other baseline methods for temporal matching partially because high temporal resolution image reconstruction using baseline methods are not sufficiently accurate (as shown in Fig.~\ref{fig:htr} and Fig.~\ref{fig:htr-curves}), resulting in inaccurate material basis images. 

\subsection{Experimental Human Subject Studies}

\begin{figure}[h]
	\centering
	\includegraphics[width=0.45\textwidth, keepaspectratio=true]{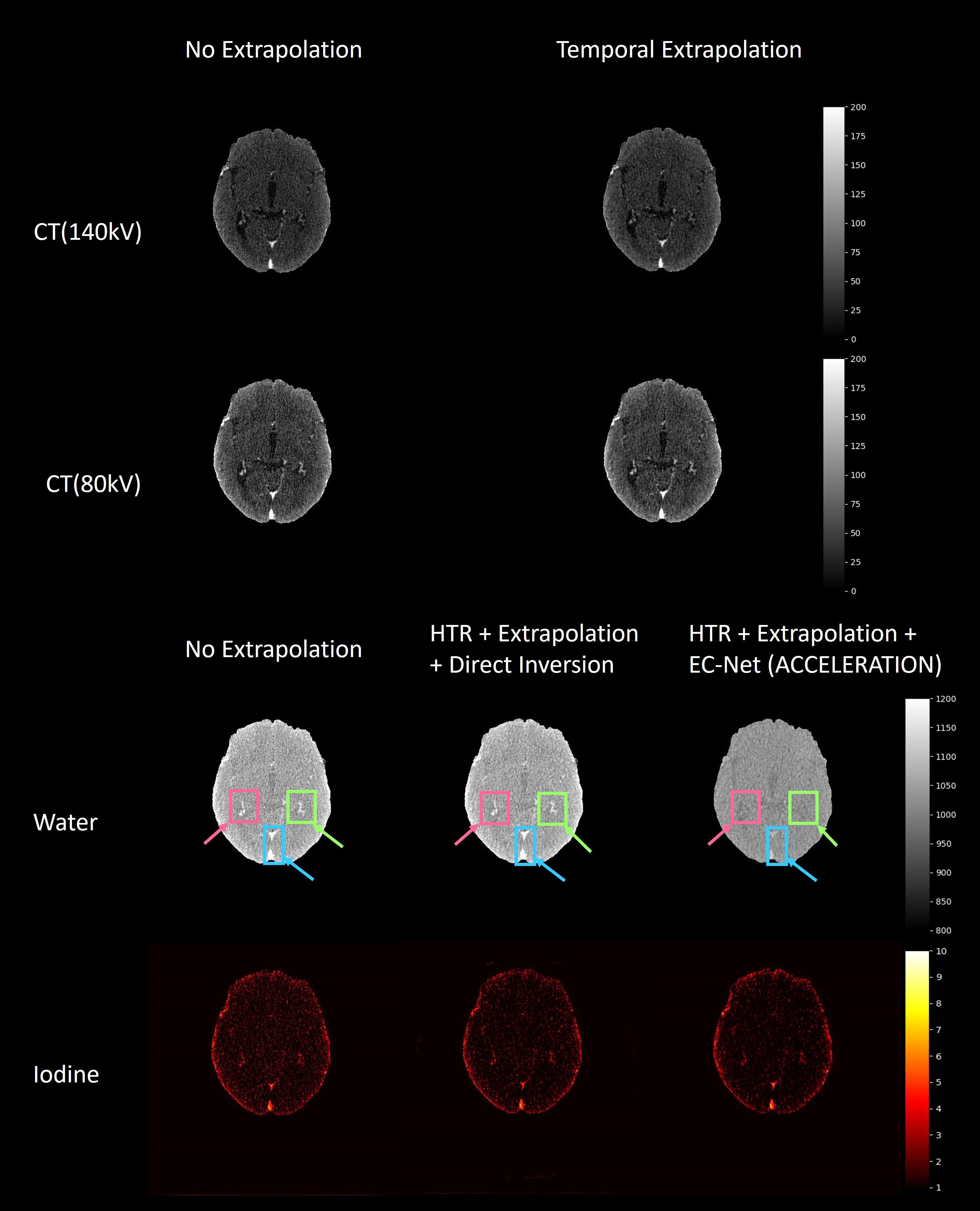}
	\caption{Material quantification results using in vivo human subject studies. The first column shows the reconstruction without temporal extrapolation, the second column shows the results using the proposed HTR and temporal extrapolation methods and using the direct inversion for material quantification. The third column shows the material quantification using the proposed ACCELERATION.}
	\label{fig:human}
\end{figure}

Results generated from human subject studies were also used to assess the overall qualitative image quality for imaged objects with complex anatomy and realistic temporal dynamics. Direct material quantification without and with HTR and temporal extrapolation were used as the baseline methods and the results were shown in the first and second columns of Fig.~\ref{fig:human}. ACCELERATION images are shown in the third row of the figure. As demonstrated by the figure, intensity values of iodinated vessels in the obtained water basis images are mitigated using ACCELERATION while other baseline methods cannot. As shown in the figure, iodinated vessels are erroneously enhanced in the water basis image using HRT and direct inversion (the second column of the figure). The image-domain direct inversion cannot produce accurate water basis image partially because the direct inversion does not incorporate the full X-ray spectrum into the physical forward model of the material decomposition process. Note that, in Fig.~\ref{fig:human}, the material basis images of baseline methods (No Extrapolation and HTR + Extrapolation + Direct Inversion) were reconstructed using in-house implementation. It is necessary to emphasize that the in-house implementation of these baseline methods is different from the reconstruction algorithms implemented in the commercial workstation provided by the vendor. 

\section{Discussion and Conclusion}

\subsection{Potential Limitations and Future Work}

This work has several potential limitations which need further investigations in future studies. 

First, there is no proprietary hyperparameter to control the trade-off between reconstruction accuracy and noise level of the reconstructed images in HTR or EC-Net. To control the trade-off, one must empirically stop the optimization process at an early stage. The immediate future work is to develop a new optimization framework to solve the optimization problem of HTR and EC-Net with theoretical convergence, explicit hyperparameters to control the trade-off between reconstruction accuracy and noise level, and other flexibilities to obtain imaging-task-specific image quality control. 

Second, the data acquisition and contrast injection protocol in the human subject study presented in the paper were not designed and optimized for time-resolved angiography imaging task. To apply ACCELERATION to this imaging task, the data acquisition and contrast injection protocols such as the injection rate, volume and concentration of contrast agent, the rate and volume of saline chasing, the timing of data acquisition, etc., need to be specifically optimized to maximally benefit from ACCELERATION for the time-resolved angiography imaging task.  

Third, the performance of HTR depends on the detailed implementation of data classification, $\{S_t\}$, and the selection of the number of time frames, $T$, in Eq.~(\ref{eq:limited-view-reconstruction}). Currently, empirical implementations of $\{S_t\}$ and $T$ have been applied in solving Eq.~(\ref{eq:limited-view-reconstruction}). The immediate future work is to systematically investigate the performance dependence of the quantitative accuracy of the reconstructed time-resolved images on different implementations of $\{S_t\}$, $T$, or the combination of both of them. 

Fourth, the sensitivity and specificity of ACCELERATION in resolving temporal dynamics induced by realistic diseases are under investigation. In our future studies, we will rely on in vivo animal studies involving stroke models with different severity levels to systematically show the potential clinical benefits of ACCELERATION.

\subsection{Conclusion}

In this work, it has been demonstrated that using ACCELERATION, a low-cost sequential-scanning data acquisition scheme to implement DECT has been achieved for contrast-enhanced imaging. Results shown in this paper demonstrated the improvement of quantification accuracy and image quality using ACCELERATION. 

\IEEEtriggeratref{70}

\bibliographystyle{IEEEtran}

\bibliography{ref}

\end{document}